\documentclass[copyright]{eptcs}  
\usepackage{epsfig, amssymb, verbatim}

\newtheorem{lemma}{Lemma}

\newtheorem{theorem}[lemma]{Theorem}
\newtheorem{proposition}[lemma]{Proposition}

\newcommand{\rlem}[1]{Lemma~\ref{#1}}

\newcommand{\rthm}[1]{Theorem~\ref{#1}}
\newcommand{\rpro}[1]{Proposition~\ref{#1}}

\newcommand{\nsc}[1]{$\mathbb{NSC}({#1})$}

\title{Nondeterministic State Complexity for
Suffix-Free Regular Languages}

\author{Yo-Sub Han\thanks{Han was supported by the
Basic Science Research Program through NRF funded by
MEST~(2010-0009168).}
\institute{Department of Computer Science, Yonsei University
Seoul 120-794, Republic of Korea}
\email{emmous@cs.yonsei.ac.kr}
\and 
Kai Salomaa\thanks{Salomaa was supported by the
Natural Sciences and Engineering Research Council of Canada
Grant~OGP0147224.}
\institute{School of Computing, Queen's University
Kingston, Ontario K7L 3N6, Canada}
\email{ksalomaa@cs.queensu.ca} 
}

\def\titlerunning{Nondeterministic State Complexity for
Suffix-Free Regular Languages}
\def\authorrunning{Y.-S. Han and K. Salomaa}

\begin{document}

\maketitle

\begin{abstract}
We investigate the nondeterministic state complexity of basic
operations for
suffix-free regular languages.
The nondeterministic state complexity of
an operation is the number of states that are
necessary and sufficient in the worst-case for a minimal
nondeterministic finite-state automaton that
accepts the language obtained from the operation.
We consider basic operations 
(catenation, union, intersection, Kleene star,  reversal and
complementation) and establish matching upper and lower bounds
for each operation. In the case of complementation the
upper and lower bounds differ by an additive constant of two.

Keywords: nondeterministic state complexity, suffix-free regular
languages, suffix codes
\end{abstract}

\section{Introduction}\label{se:intro}
Codes are useful in information processing, data compression, 
cryptography and information transmission~\cite{JurgensenK97}. 
Some well-known examples are prefix codes, suffix codes, bifix codes
and infix codes. People use different codes for different application
domains based on the characteristic of each 
code~\cite{BerstelP85,JurgensenK97}.
Since a code is a {\em language}, the conditions that classify
codes define subfamilies of language families.
For regular languages, for example, the prefix-freeness of prefix
codes defines the family of prefix-free regular languages, 
which is a proper subfamily of regular languages.
Prefix-freeness is fundamental in coding theory; for
example, Huffman codes are prefix-free sets. The advantage of
prefix-free codes
is that we can decode a given encoded string deterministically.
The symmetric to prefix codes are suffix codes; given a prefix
code, its reversal is always a suffix code. 
However, suffix codes have their own unique characteristics and are not
always completely symmetric to prefix codes. For instance, a
finite-state automaton~(FA) is prefix-free if and only if it has no
out-transitions from any final state. 
If we think of a reversal of this FA, we can think of an FA whose
start state has no in-transitions. 
However, this condition is just a necessary condition for being 
suffix-free but not sufficient.
Thus, we often need to examine the suffix-free case separately.

Regular languages are given by FAs or regular
expressions. There are two main types of FAs:  deterministic
finite-state
automata~(DFAs) and  nondeterministic finite-state
automata~(NFAs). NFAs provide exponential savings in space compared
with
DFAs but the problem to convert a given DFA to an equivalent minimal
NFA is PSPACE-complete~\cite{JiangR93}.
For finite languages, Salomaa and Yu~\cite{SalomaaY98} showed that
$O(k^{\frac{n}{\log_2 {k+1}}})$ is a tight bound for converting an
$n$-state NFA to a DFA, where $k$ is the size of an input alphabet.

There are at least two different models for
the state complexity of operations:
The deterministic state complexity model considers minimal DFAs
and the nondeterministic state complexity considers minimal NFAs.

Yu et al.~\cite{Yu01,YuZS94} investigated
the deterministic
state complexity for various operations on regular languages.
As special cases of state complexity,
C{\^a}mpeanu et al.~\cite{CampeanuCSY01} and Han and
Salomaa~\cite{HanS08} examined
the deterministic state complexity of finite languages.
Pighizzini and Shallit~\cite{PighizziniS02} investigated the
deterministic state complexity of unary language operations.
Moreover, Han et al.~\cite{HanSW09} studied the deterministic state
complexity of prefix-free regular languages and Han and
Salomaa~\cite{HanS09} looked into the deterministic state complexity
of suffix-free regular languages.
After writing this paper, we have found out that 
Jir{\'a}skov{\'a} and Olej{\'a}r~\cite{JiraskovaO09}
have also considered the nondeterministic state
complexity of union and intersection for suffix-free languages.
They have established a tight bound for union and intersection
using binary languages.
There are several
other results with respect to the state complexity of various
operations~\cite{CampeanuSY02,Domaratzki02,SalomaaWY04}.

Holzer and Kutrib~\cite{HolzerK03} studied the nondeterministic
state complexity of regular languages.
Jir{\'a}sek et
al.~\cite{JirasekJS05} examined the nondeterministic
state complexity of complementation of regular languages.
Recently, Han et al.~\cite{HanSW09b} investigated the nondeterministic
state complexity of prefix-free regular languages. 
As a continuation of our research for the operational nondeterministic
state complexity of subfamilies of regular languages, we consider 
the nondeterministic state complexity of suffix-free regular languages. 
Since suffix codes are one of the fundamental classes of codes, 
it is important to calculate the precise bounds.
Moreover, determining the state complexity of operations 
on fundamental subfamilies of
the regular languages can provide valuable insights on connections
between
restrictions placed on language definitions and descriptional
complexity.

In Section~\ref{se:pre}, we define some basic notions.
In Section~\ref{se:complexity}, we examine the worst-case
nondeterministic state complexity of
basic operations (union, catenation, intersection, Kleene star,
reversal and complementation) of
suffix-free regular languages.
Except for the complementation operation, we prove that the 
results are tight by giving general lower
bound examples that match the upper bounds.

We give a comparison table between the deterministic state complexity
and the nondeterministic state complexity in Section~\ref{se:con}.

\section{Preliminaries}\label{se:pre}
Let $\Sigma$ denote a finite alphabet of characters
and $\Sigma^*$ denote the set of all
strings over $\Sigma$. The size~$|\Sigma|$ of $\Sigma$ is
the number of characters in $\Sigma$.
A language over $\Sigma$ is any subset of $\Sigma^*$.
The symbol~$\emptyset$ denotes the empty language and the
symbol~$\lambda$ denotes the null string.
For strings~$x,y$ and $z$,
we say that $x$ is a {\em suffix\/} of $y$ if $y = zx$.
We define a (regular) language~$L$ to be suffix-free if a string~$x\in
L$ is not a suffix of any other strings in $L$.
Given a string~$x$ in a set~$X$ of strings,
let $x^R$ be the reversal of $x$, in which case $X^R = \{x^R \mid x\in
X\}$.

An FA~$A$ is specified by a tuple
$(Q,\Sigma,\delta,s,F)$, where $Q$ is a finite set of states,
$\Sigma$ is an input alphabet,
$\delta: Q\times \Sigma \to 2^Q$ is a transition function,
$s\in Q$ is the start state and $F\subseteq Q$ is a set of final
states.
If $F$ consists of a single state~$f$, then we use $f$ instead of
$\{f\}$
for simplicity.
Let $|Q|$ be the number of states in $Q$.
We define the size~$|A|$ of $A$ to be
the number of states in $A$; namely $|A| = |Q|$.
For a transition~$q \in \delta(p,a)$ in $A$,
we say that $p$ has an {\em out-transition\/} and $q$ has an {\em
in-transition\/}. Furthermore, $p$ is a {\em source state\/} of $q$
and
$q$ is a {\em target state\/} of $p$.
We say that $A$ is {\em non-returning\/} if the start state of $A$
does
not
have any in-transitions and $A$ is {\em non-exiting\/} if all final
states of $A$ do not have any out-transitions.
If $\delta(q, a)$ has a single element~$q'$, then
we denote $\delta(q, a) = q'$ instead of $\delta(q, a) = \{q'\}$
for simplicity.

A string~$x$ over $\Sigma$ is accepted by~$A$ if there is a labeled
path from~$s$ to a final state such that this path spells out $x$.
We call this path an {\em accepting path}.
Then, the language~$L(A)$ of $A$ is the set
of all strings spelled out by accepting paths in $A$.
We say that a state of $A$ is {\em useful\/} if it appears in an
accepting path in $A$; otherwise, it is {\em useless\/}.
Unless otherwise mentioned, in the following we assume that all states
of an FA are useful.

We say that an FA~$A$ is a suffix-free FA if $L(A)$ is suffix-free. 
Notice that a suffix-free FA must be non-returning by definition.
We assume that a given NFA has no $\lambda$-transitions since we can
always transform
an $n$-state NFA with $\lambda$-transitions to an equivalent
$n$-state NFA without $\lambda$-transitions~\cite{HopcroftU79}.

For complete background knowledge in automata theory, the reader may
refer to textbooks~\cite{HopcroftU79,Shallit08,Wood87}.

Before tackling the problem,
we present a nice technique that gives a lower bound for the
size of NFAs and
establish
a lemma that is crucial to prove the tight bound for the
nondeterministic state complexity in the following sections.
Notice that an FA for a non-trivial suffix-free regular language~$L$
(namely, $L \ne \{\lambda\}$) must have at least 2~states
since such FA needs at
least one start state and one final state.

\begin{proposition}[(The fooling set technique~\cite{Birget92,GlaisterS96})]\label{pro:GS}
Let $L \subseteq \Sigma^*$ be a regular language. Suppose that there
exists a set of pairs
$$
P = \{ (x_i, w_i) \mid 1 \le i \le n \} 
$$
such that 
\begin{enumerate}
\item For all $i$ with $1 \le i \le n$, we have $x_i w_i \in L$;
\item For all $i, j$ with $1 \le i, j \le n$ and $i \ne j$, at least
one of $x_i w_j \notin L$ and $x_jw_i \notin L$ holds.
\end{enumerate}

Then, a minimal NFA for $L$ has at least $n$~states.
\end{proposition}

The set~$P$ satisfying the conditions of \rpro{pro:GS} is called a
{\em fooling set\/} for $L$.
The fooling set technique was first proposed by Birget~\cite{Birget92}.
A related technique was considered by Glaister and
Shallit~\cite{GlaisterS96}.

\begin{lemma}\label{lem:min_nfa}
Let $n \ge 2$ be an arbitrary integer. 
A minimal NFA of the suffix-free language~$L_1 = L(b(a^{n-1})^*)$ with
$n \ge 2$ or
of the suffix-free language~$L_2 = L(b(a^{n-2})^*b)$ with $n \ge 3$
has $n$~states. 
\end{lemma}

We use \nsc{L} to denote the number of states of a minimal NFA for
$L$; namely, \nsc{L} is the 
nondeterministic state complexity of $L$.

\section{State Complexity}\label{se:complexity}

We first examine the nondeterministic state complexity of binary
operations (union, catenation and intersection) 
for suffix-free regular languages. Then, we study the unary
operation cases (Kleene star, reversal and complementation). 
We rely on a unique structural
property of a suffix-free FA for obtaining upper bounds: 
The start state does not have any
in-transitions (the non-returning property).

\subsection{Union}
Han and Salomaa~\cite{HanS09} showed that $mn - (m+n) +2$ is the state 
complexity of the union of an $m$-state suffix-free DFA and an 
$n$-state suffix-free DFA using the Cartesian product of states. For 
the NFA state complexity, we directly construct an NFA for the union
of two suffix-free regular languages without the Cartesian product. 
The construction relies on nondeterminism and the fact that
the computation of a suffix-free FA cannot return to the start state.

\begin{theorem}\label{thm:union}
Given two suffix-free regular languages~$L_1$ and $L_2$, the nondeterministic
state complexity~\nsc{L_1\cup L_2} for $L_1\cup L_2$ is $m+n-1$, where
$m$ = \nsc{L_1}, $n$ = \nsc{L_2}, $m, n \ge 2$ and $|\Sigma| \ge 2$.
\end{theorem}

\subsection{Catenation}
For the catenation operation,  
$(2m-1)2^{n-1}$ is the state complexity for the DFA case~\cite{YuZS94}
and $m+n$ is the state complexity for the NFA case~\cite{HolzerK03}. 
Thus, there is an exponential gap between two cases. For the
prefix-free regular languages, the state complexity is linear in the
sizes of the component automata in both DFA and NFA cases because of a
unique structural property of a prefix-free automaton~\cite{HanSW09b}.
The deterministic state complexity of the catenation of suffix-free
regular languages is $(m-1)2^{n-2}+1$~\cite{HanS09}.

\begin{theorem}\label{thm:caten}
Given two suffix-free regular languages~$L_1$ and $L_2$, 
the nondeterministic state complexity~\nsc{L_1L_2} for $L_1L_2$ 
is $m+n-1$, 
where $m$ = \nsc{L_1}  and $n$ = \nsc{L_2}.
\end{theorem}

\subsection{Intersection}
Given two FAs~$A = (Q_1, \Sigma, \delta_1, s_1, F_1)$ and $B = (Q_2,
\Sigma, \delta_2, s_2, F_2)$, we can construct an 
FA~$M =  (Q_1 \times Q_2, \Sigma, \delta, (s_1, s_2), F_1 \times
F_2)$
for the intersection of $L(A)$ and $L(B)$ based on the Cartesian
product of states, where
\[
\delta((p,q), a) = (\delta_1(p,a), \delta_2(q,a)) \textrm{~for~} p \in
Q_1, q \in Q_2 \textrm{~and~} a \in \Sigma.
\]

From the Cartesian product, we know that the upper bound for the
intersection of two FAs is at most $mn$, where $m$ and $n$ are the
numbers of states for $A$ and $B$. We now examine $M$ and reduce the
upper bound based on the suffix-freeness of input FAs. 
Let $A$ and $B$ be suffix-free. This implies that both $A$ and $B$
are non-returning and, thus, $s_1$ and $s_2$ do not have any
in-transitions. 

\begin{proposition}\label{pro:inter}
All states~$(s_1, q)$ and $(p, s_2)$, for $p(\ne s_1) \in Q_1$ and 
$q(\ne s_2) \in Q_2$, are unreachable from $(s_1, s_2)$ in $M$
since $L(A)$ and $L(B)$ are suffix-free.
\end{proposition}

Based on \rpro{pro:inter}, we remove all unreachable states and reduce
the upper bound as follows:
$$
mn - (m-1) - (n-1)  = mn - (m+n) + 2.
$$

Namely, $ mn - (m+n) + 2 $~states are sufficient for $L(A)\cap L(B)$
when both $A$ and $B$ are non-returning.

\begin{theorem}\label{thm:inter}
Given two suffix-free regular languages~$L_1$ and $L_2$, the
nondeterministic state complexity~\nsc{L_1\cap L_2} for $L_1\cap L_2$
is $mn - (m+n) +2$, where 
$m$ = \nsc{L_1}, $n$ = \nsc{L_2} and $|\Sigma| \ge 3$.
\end{theorem}

\rthm{thm:inter} considers when \nsc{L_1}, \nsc{L_2} $\ge 2$. 
If either of them is
1, then \nsc{L_1\cap L_2} = 1 since the single state suffix-free regular
language is $\{\lambda\}$.
The deterministic state complexity of the intersection of two
suffix-free DFAs is $mn-2(m+n)+6$~\cite{HanS09}. The complexity gap 
between the DFA case and the NFA case is because of the sink
state. An NFA does not need to have a sink state.

\subsection{Kleene Star}
We examine the Kleene star operation of suffix-free NFAs. Han and
Salomaa~\cite{HanS09} investigated the deterministic state complexity
for Kleene star and demonstrated that
$2^{m-2} +1$~states are necessary and sufficient in the worst-case for
an $m$-state suffix-free DFA.

\begin{theorem}\label{thm:kplus}
Given a suffix-free regular language~$L$, the nondeterministic state
complexity~\nsc{L^*} for $L^*$ is $m$, where $m$ = \nsc{L}.
\end{theorem}

\subsection{Reversal}
Given an $m$-state NFA~$A$, \nsc{L(A)} is in general
$m+1$~\cite{HolzerK02}. If $L(A)$ is prefix-free, then we know that
\nsc{L(A)} is $m$~\cite{HanSW09b}.  

The upper bound~$m+1$ is based on the simple NFA construction for
$L^R$ from $A$ for $L$: 
We flip the transition directions and make the start state to be
a final state and all final states to be start states of $A$. 
Now we have an NFA with multiple start states. 
We introduce a new start state and make a
$\lambda$-transition from the new start state to the original start
states. Then, we apply the $\lambda$-transition removal
technique~\cite{HopcroftU79}, which does not change the number of states. 
Thus, we have an $m+1$-state NFA for $L^R$.

Now we consider a lower bound for reversal.
It seems difficult to apply the fooling set method
for this operation. For the below lemma we use an
ad hoc proof that has been modified from the corresponding
argument used in Holzer and Kutrib~\cite{HolzerK03} for the reversal of general
regular languages.

\begin{lemma}
\label{revlem}
Let $\Sigma = \{ a, b, c, d \}$ and $m \geq 4$. 
There exists a suffix-free regular language over $\Sigma$
with ${\mathbb{NSC}}(L) \leq m$ such that
${\mathbb{NSC}}(L^R) = m+1$.
\end{lemma}

In the construction used for Lemma~\ref{revlem},
when $m \geq 4$ the symbol
$d$ can be replaced by $b$ or $c$. We have stated the
construction using a four-letter alphabet for the sake
of easier readability. We do not know whether the lower
bound $m+1$ can be reached by the reversal of suffix-free
regular languages over a two-letter alphabet.

Using the general upper bound from Holzer and Kutrib~\cite{HolzerK03},
Lemma~\ref{revlem} gives the following statement:

\begin{theorem}
If $L$ is a suffix-free regular language recognized
by an NFA with $m$ states, then ${\mathbb{NSC}}(L^R) \leq m+1$.
The bound $m+1$ can be reached by suffix-free languages over a
three letter alphabet when $m \geq 4$\footnote{An anonymous referee of
the paper has suggested a 
different lower bound construction over a 3-letter alphabet
that works also in the case $m=3$.}.
\end{theorem}

\subsection{Complementation of suffix-free regular
languages}\label{se:complement}

The complementation of NFA is an expensive operation with respect to
state complexity. Meyer and Fischer~\cite{MeyerF71} already noticed
that the transforming an $m$-state NFA to a DFA requires $2^m$~states. 
The complementation of an $m$-state DFA does not require additional
states since it simply interchanges final states and non-final states.
Thus, based on the subset construction, we know that $2^m$~states are
sufficient for the complementation of an $m$-state NFA. 
Jir\'{a}skov\'{a}~\cite{Jiraskova05} showed that $2^m$ states are
necessary for the tight bound when $|\Sigma| = 2$. 

\begin{lemma}\label{lem:complement_upper}
Given an $m$-state suffix-free NFA~$A = (Q, \Sigma, \delta, s, F)$, 
$2^{m-1}+1$~states are
sufficient for its complementation language~$\overline{L(A)}$.
\end{lemma}

\begin{lemma}
\label{nlem1}
Let $\Sigma = \{ a, b, c \}$  and 
$L_1 \subseteq \{ a, b \}^*$ be a regular language.
Let $L \subseteq \Sigma^*$ be a regular language
such that
\begin{equation}
\label{ketta2}
L \cap (c \cdot \Sigma^*) = c \cdot L_1.
\end{equation}
Then ${\mathbb{NSC}}(L) \geq {\mathbb{NSC}}(L_1) - 1$.
\end{lemma}

If in the statement of Lemma~\ref{nlem1} the
language  $L$ is suffix-free, the proof
implies that ${\mathbb{NSC}}(L) \geq {\mathbb{NSC}}(L_1)$. In this
case, the constructed NFA $B$ does need the start state
of the original NFA $A$ since
$A$ is non-returning. However, below Lemma~\ref{nlem1}
will be used  for a complementation of  suffix-free
languages (that need not be suffix-free)
and the bound cannot be improved in this way.

\begin{lemma} \label{nlem2}
Let $\Sigma = \{ a, b, c \}$ and $m \geq 2$. There exists
a suffix-free regular language $L \subseteq \Sigma^*$ such that
$$
{\mathbb{NSC}}(L) \leq m \mbox{ and } {\mathbb{NSC}}(\overline{L}) \geq
2^{m-1} - 1.
$$
\end{lemma}

The results of \rlem{lem:complement_upper}
and Lemma~\ref{nlem2} give the following.

\begin{theorem}
\label{nth1}
Given a suffix-free regular
language $L$ having an NFA with $m$ states,
${\mathbb{NSC}}(\overline{L}) \leq 2^{m-1}+1$. There exists
a  suffix-free regular language $L$ over a three letter alphabet
such that ${\mathbb{NSC}}(L) = m$
and ${\mathbb{NSC}}(\overline{L}) \geq 2^{m-1} - 1$.
\end{theorem}

Theorem~\ref{nth1} gives the precise worst-case
nondeterministic state complexity of complementation within
a constant of two. The worst-case example for
complementation in Jir\'{a}skov\'{a}~\cite{Jiraskova05}
uses a binary alphabet, however, our construction needs an
additional symbol to make the languages suffix-free. We do not
know what is the nondeterministic state complexity of complementation
for suffix-free languages over a binary alphabet.

\section{Conclusions}\label{se:con}
We have investigated the nondeterministic
state complexity of basic operations for
suffix-free regular languages. We have relied on a unique structural
property of a suffix-free FA: The start state does not have any
in-transitions.
Based on this property,
we have examined the nondeterministic state complexity with respect to
catenation, union, intersection, Kleene star, reversal and
complementation.
Table~\ref{fi:summary} shows the comparison between the deterministic
state complexity and the nondeterministic the state complexity.

\begin{table}[htb]
\begin{center}
\begin{tabular}{ l   l l }
operation & suffix-free DFAs & suffix-free NFAs\\
\hline\hline
$L_1 \cdot L_2$ & $(m-1)2^{n-2}+1$ & $ m+n-1$\\
$L_1 \cup L_2$ & $mn - (m+n)+2$ & $m+n-1 $\\
$L_1 \cap L_2$  & $mn - 2(m+n)+6$ & $ mn-2(m+n)+2$\\
$L_1^*$ &  $2^{m-2}+1$ & $m $\\
$L_1^R$ &  $2^{m-2}+1$ & $m+1 $\\
$\overline{L_1}$ & $m$ & $2^{m-1}\pm 1$\\
\end{tabular}
\end{center}
\caption{State complexity of basic operations between suffix-free
DFAs and NFAs.}
\label{fi:summary}
\end{table}

\section*{Acknowledgements}
We wish to thank the referees for providing us with constructive
comments and suggestions. 

\bibliographystyle{abbrv}
\bibliography{ref}

\begin{thebibliography}{10}

\bibitem{BerstelP85}
J.~Berstel and D.~Perrin.
\newblock {\em Theory of Codes}.
\newblock Academic Press, Inc., 1985.

\bibitem{Birget92}
J.-C. Birget.
\newblock Intersection and union of regular languages and state complexity.
\newblock {\em Information Processing Letters}, 43(4):185--190, 1992.

\bibitem{CampeanuCSY01}
C.~C{\^a}mpeanu, K.~{Culik~II}, K.~Salomaa, and S.~Yu.
\newblock State complexity of basic operations on finite languages.
\newblock In {\em Proceedings of WIA'99}, Lecture Notes in Computer Science
  2214,  60--70, 2001.

\bibitem{CampeanuSY02}
C.~C{\^a}mpeanu, K.~Salomaa, and S.~Yu.
\newblock Tight lower bound for the state complexity of shuffle of regular
  languages.
\newblock {\em Journal of Automata, Languages and Combinatorics},
  7(3):303--310, 2002.

\bibitem{Domaratzki02}
M.~Domaratzki.
\newblock State complexity of proportional removals.
\newblock {\em Journal of Automata, Languages and Combinatorics},
  7(4):455--468, 2002.

\bibitem{GlaisterS96}
I.~Glaister and J.~Shallit.
\newblock A lower bound technique for the size of nondeterministic finite
  automata.
\newblock {\em Information Processing Letters}, 59(2):75--77, 1996.

\bibitem{HanS08}
Y.-S. Han and K.~Salomaa.
\newblock State complexity of union and intersection of finite languages.
\newblock {\em International Journal of Foundations of Computer Science},
  19(3):581--595, 2008.

\bibitem{HanS09}
Y.-S. Han and K.~Salomaa.
\newblock State complexity of basic operations on suffix-free regular
  languages.
\newblock {\em Theoretical Computer Science}, 410(27-29):2537--2548, 2009.

\bibitem{HanSW09b}
Y.-S. Han, K.~Salomaa, and D.~Wood.
\newblock Nondeterministic state complexity of basic operations for prefix-free
  regular languages.
\newblock {\em Fundamenta Informaticae}, 90(1-2):93--106, 2009.

\bibitem{HanSW09}
Y.-S. Han, K.~Salomaa, and D.~Wood.
\newblock Operational state complexity of prefix-free regular languages.
\newblock In {\em Automata, Formal Languages, and Related Topics - Dedicated to
  Ferenc G{\'e}cseg on the occasion of his 70th birthday},  99--115, 2009.

\bibitem{HolzerK02}
M.~Holzer and M.~Kutrib.
\newblock Unary language operations and their nondeterministic state
  complexity.
\newblock In {\em Proceedings of DLT'02}, Lecture Notes in Computer Science
  2450,  162--172, 2002.

\bibitem{HolzerK03}
M.~Holzer and M.~Kutrib.
\newblock Nondeterministic descriptional complexity of regular languages.
\newblock {\em International Journal of Foundations of Computer Science},
  14(6):1087--1102, 2003.

\bibitem{HopcroftU79}
J.~Hopcroft and J.~Ullman.
\newblock {\em Introduction to Automata Theory, Languages, and Computation}.
\newblock Addison-Wesley, Reading, MA, 2 edition, 1979.

\bibitem{JiangR93}
T.~Jiang and B.~Ravikumar.
\newblock Minimal {NFA} problems are hard.
\newblock {\em SIAM Journal on Computing}, 22(6):1117--1141, 1993.

\bibitem{JirasekJS05}
J.~Jir{\'a}sek, G.~Jir{\'a}skov{\'a}, and A.~Szabari.
\newblock State complexity of concatenation and complementation.
\newblock {\em International Journal of Foundations of Computer Science},
  16(3):511--529, 2005.

\bibitem{Jiraskova05}
G.~Jir\'{a}skov\'{a}.
\newblock State complexity of some operations on binary regular languages.
\newblock {\em Theoretical Computer Science}, 330(2):287--298, 2005.

\bibitem{JiraskovaO09}
G.~Jir{\'a}skov{\'a} and P.~Olej{\'a}r.
\newblock State complexity of union and intersection of binary suffix-free
  languages.
\newblock In {\em Proceedings of Workshop on Non-Classical Models for Automata
  and Applications},  151--166, 2009.

\bibitem{JurgensenK97}
H.~J\"urgensen and S.~Konstantinidis.
\newblock Codes.
\newblock In G.~Rozenberg and A.~Salomaa, editors, {\em Word, Language,
  Grammar}, volume~1 of {\em Handbook of Formal Languages},  511--607.
  Springer-Verlag, 1997.

\bibitem{MeyerF71}
A.~R. Meyer and M.~J. Fischer.
\newblock Economy of description by automata, grammars, and formal systems.
\newblock In {\em Proceedings of the Twelfth Annual IEEE Symposium on Switching
  and Automata Theory},  188--191, 1971.

\bibitem{PighizziniS02}
G.~Pighizzini and J.~Shallit.
\newblock Unary language operations, state complexity and {J}acobsthal's
  function.
\newblock {\em International Journal of Foundations of Computer Science},
  13(1):145--159, 2002.

\bibitem{SalomaaWY04}
A.~Salomaa, D.~Wood, and S.~Yu.
\newblock On the state complexity of reversals of regular languages.
\newblock {\em Theoretical Computer Science}, 320(2-3):315--329, 2004.

\bibitem{SalomaaY98}
K.~Salomaa and S.~Yu.
\newblock {NFA} to {DFA} transformation for finite languages over arbitrary
  alphabets.
\newblock {\em Journal of Automata, Languages and Combinatorics},
  2(3):177--186, 1998.

\bibitem{Shallit08}
J.~Shallit.
\newblock {\em A Second Course in Formal Languages and Automata Theory}.
\newblock Cambridge University Press, New York, NY, USA, 2008.

\bibitem{Wood87}
D.~Wood.
\newblock {\em Theory of Computation}.
\newblock John Wiley \& Sons, Inc., New York, NY, 1987.

\bibitem{Yu01}
S.~Yu.
\newblock State complexity of regular languages.
\newblock {\em Journal of Automata, Languages and Combinatorics},
  6(2):221--234, 2001.

\bibitem{YuZS94}
S.~Yu, Q.~Zhuang, and K.~Salomaa.
\newblock The state complexities of some basic operations on regular languages.
\newblock {\em Theoretical Computer Science}, 125(2):315--328, 1994.

\end{thebibliography}
\end{document}